\pdfoutput=1

\documentclass[11pt]{article}

\usepackage[preprint]{acl/acl}

\usepackage[shortlabels]{enumitem}

\usepackage{times}
\usepackage{latexsym}

\usepackage[T1]{fontenc}

\usepackage[utf8]{inputenc}

\usepackage{microtype}

\usepackage{inconsolata}

\usepackage{graphicx}

\usepackage{caption}
\usepackage{amsthm}

\usepackage{subcaption}
\usepackage{graphicx}

\usepackage{natbib}

\usepackage{amsfonts,amsmath,amssymb}
\usepackage{bbm}
\usepackage{xspace}

\usepackage{multicol}
\usepackage{enumitem}

\usepackage{algorithm}
\usepackage{algpseudocode}

\usepackage{wrapfig}

\title{Investigating social alignment via mirroring in a system of interacting language models}

\author{Harvey McGuinness$^{*\dagger}$ \\
Johns Hopkins University \\
\And
Tianyu Wang$^{*}$ \\
Johns Hopkins University
\And
Carey E. Priebe \\
Johns Hopkins University \AND
Hayden Helm \\
Helivan Research
}

\newcommand\blfootnote[1]{%
  \begingroup
  \renewcommand\thefootnote{}\footnote{#1}%
  \addtocounter{footnote}{-1}%
  \endgroup
}

\begin{document}
\maketitle

\blfootnote{\\
$^{*} $ denotes equal contribution. \\
$ ^{\dagger} $ corresponding author: \texttt{hmcguin1@jhu.edu}.
}

\begin{abstract}
Alignment is a social phenomenon wherein individuals share a common goal or perspective.
Mirroring, or mimicking the behaviors and opinions of another individual, is one mechanism by which individuals can become aligned.
Large scale investigations of the effect of mirroring on alignment have been limited due to the scalability of traditional experimental designs in sociology.
In this paper, we introduce a simple computational framework that enables studying the effect of mirroring behavior on alignment in multi-agent systems.
We simulate systems of interacting large language models in this framework and characterize overall system behavior and alignment with quantitative measures of agent dynamics.
We find that system behavior is strongly influenced by the range of communication of each agent and that these effects are exacerbated by increased rates of mirroring.
We discuss the observed simulated system behavior in the context of known human social dynamics.
\end{abstract}

A key phenomenon underlying the formation of human groups is social alignment \cite{Ransom}. 
This phenomenon, through which individuals adopt the behaviors of the individuals they interact with, has been shown to support the creation of social relationships through the promotion of positive commonality \cite{ACM}. 
In short, like attracts like, and it is through social alignment that individuals behave more alike one another in order to foster cohesive relationships. 

One of the most direct mechanisms to achieve social alignment is social mirroring \cite{Hasson_Frith_2016a}. This occurs primarily subconsciously, as individuals will mimic (or mirror) the facial expressions, posture, mannerisms, and other behaviors of those with whom they interact \cite{bahar}. However, social mirroring is not limited to physical mimicry, as it can also include a more complete degree of copying wherein individuals parrot speech patterns, opinions, and perspectives of those they are mirroring to appear maximally agreeable  \cite{Byrne_2005}. 

While social mirroring has been studied extensively in small group conditions -- primarily with respect to one-on-one interacting pairs \cite{galloti} -- little research has been done exploring its system-level consequences, primarily due to the infeasibility of large-scale experiments. Instead, system-level social science research has been predominantly concerned with the broad consequences of social alignment in group performance once a group has already been established \cite{local_alignment}, rather than the decentralized processes from which those groups can emerge. 


Previous theoretical work exploring foundational human social behaviors has been conducted through either purely mathematical means such as the long-tested voter model of population dynamics \cite{redner} or via overly simple computational frameworks \cite{doi:10.1073/pnas.2106292118}. While these approaches can produce excellent insights into the general social dynamics of a population, recent advances in machine learning techniques have enabled explorations of classic economic, psycholinguistic, and social psychology experiments. 
For example, the recent cohort of Large Language Models (LLMs), such as OpenAI's GPT4, are able to replicate the Ultimatum Game, Garden Path Sentences, and Milgram Shock Experiment \citep{aher}. 

Although recent LLMs are able to generate human-like outputs \citep{webb, helm2023statisticalturingtestgenerative} and replicate classic experiments without explicit training (see, e.g., \cite{webb}), human-like behavior is not sufficient to model individuals in a social system.
Indeed, it is necessary to have generative variation across agents that effectively mimics cognitive differences across people.
Prior research has shown that retrieval augmented generation (``RAG") improves the performance of LLMs by allowing them to retrieve documents that are relevant to a query from a domain-specific or knowledge-intensive database \citep{rag}. 
For social experiments, a RAG database can be considered the knowledge base or memory of an individual in the social system: the LLM will generate responses based on the combination of its pre-training data and the RAG database, just as people express their thoughts based on a history of previous experiences and the current environmental context.
Previous work studying LLM agents, such as \cite{park2023generative}, have shown that pairing a single LLM with a RAG database-per-individual can effectively promote generation diversity.

In this paper we present a computational framework to study the macroscopic effects of social alignment through social mirroring. Our framework is parameterized by important characteristics of interactions in a social system, such as the likelihood of an individual being mirrored and the range of individuals with whom someone may interact.
The framework is general to collections of agents that have the capacity to exchange information, though our analysis focuses on systems of interacting LLMs due to their ability to exchange information in natural language at a human-like capacity.

We first introduce our experiment of interacting LLMs to study the effect of mirroring on social alignment in detail. 
We then present and discuss the results before interpreting them from the perspective of comparable social science case studies.

\begin{figure*}[t]
\centering
    \includegraphics[width=\linewidth]{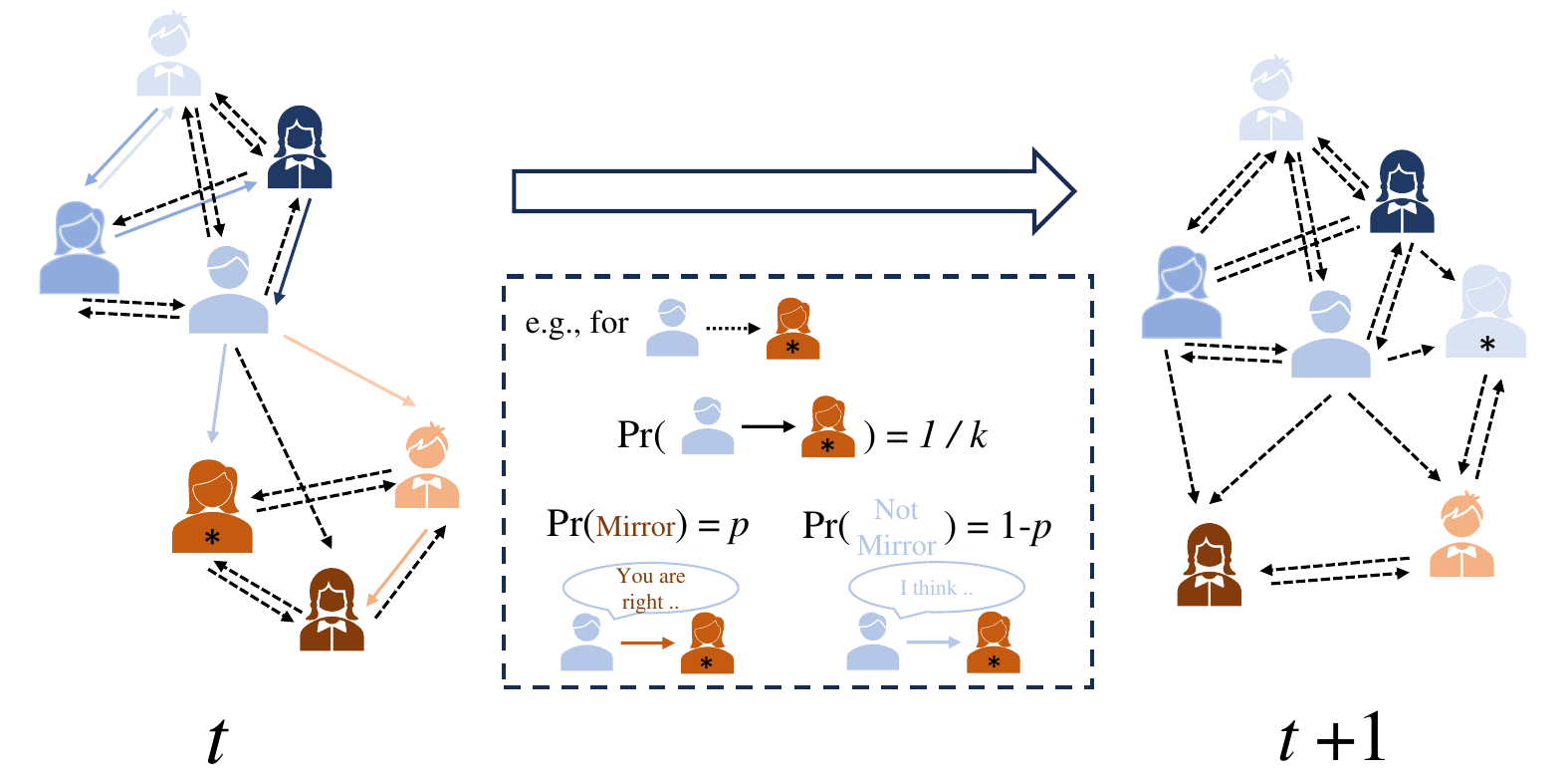}
    \caption{Illustration of simulated system dynamics with $ n = 7 $ agents with a communication range of $ k = 3 $. 
    Dashed arrows connect agents in the communication range of the receiving agent. 
    Solid arrows connect the agent that interacted with the receiving agent.
    Arrow color represents the type of interaction -- if it is the color of the receiving agent then communicating agent is mirroring, otherwise they are not mirroring.
    Mirroring occurs with probability $ p $ for each interaction.
    The opinion of an agent and the agents in its range of communication may change at time $ t + 1 $ based off its interaction at time $ t $.
    For example, the blue arrow from the light-blue male to the orange female at $ t $ affects the opinion of the orange female at $ t + 1 $ -- she becomes light blue and the agents in her communication range change.
    See Section \ref{sec:mechanics} for further details.}
\label{fig:system_illustration}
\end{figure*}
\section{Experimental Design}
\label{sec:mechanics}

We let $A_1,\hdots,A_n$ be $ n $ agents.
Each agent is Meta's \texttt{LLaMA-2-7B-Chat} \citep{touvron2023llama} equipped with a different external database.
As mentioned above, the external databases are a proxy for different knowledge bases and memories of individuals in a community. 
In our experiment, each database contains a fixed number of sentences about flowers.
These sentences are randomly sampled from a collection of sentences originally generated by prompting ChatGPT with ``Provide sentences that describe the beauty of various kinds of flowers. Sentences should not exceed 25 words.'' By limiting the size of database and the length of sentences, we control the initialization consistency across different agents. 
In our experiment each agent has a unique initialization to properly represent natural diversity in a community.

We simplify the set of possible sequences of social interactions by only allowing agent interactions at discrete time steps $ t = 1, \hdots, T $. 
The experimental protocol is composed of a measurement step and an interaction step:
\begin{itemize}
\item \textbf{(Measurement Step)} At each time $t$, we ask each agent ``Describe the prettiest flower in a single sentence.''.
We denote the answer as $ B_{i}^{(t)} $ ($ i = 1, \;\hdots, \;n;\; t = 0, \;\hdots, \;T $).
The answers will be used to measure agent alignment.
 
We use the open source embedding model \texttt{nomic-embed-v1.5} \citep{nomic} to quantitatively measure differences amongst the $ B_{i}^{(t)} $.
The embedding model maps each sentence $B_{i}^{(t)}$ to a vector $X_{i}^{(t)}\in\mathbb{R}^{768}$. 
A smaller distance between $X_{i}^{(t)}$ and $X_{j}^{(t')}$ means that $B_{i}^{(t)}$ and $B_{j}^{(t')}$ are more semantically similar and, hence, more aligned.
We capture the pairwise alignment with $ D^{(t)} $:
\begin{equation*}
(D^{(t)})_{ij}:=||X_{i}^{(t)}-X_{j}^{(t)}||_2,
\end{equation*} where $||\cdot||_2$ denotes the vector $l_2$-norm. 
We use $l_2$-norm, as opposed to cosine (dis)similarity, because we expect to extend our framework to the scenario where multiple queries are asked. 
In the multi-query scenario, $X_i^{(t)}$ becomes a matrix instead of a vector, and 
the Frobenius norm, i.e. $||\cdot||_{2,2}$, is a common choice of metric.
 


\begin{figure*}[t]
\centering
\includegraphics[width=11.4cm]{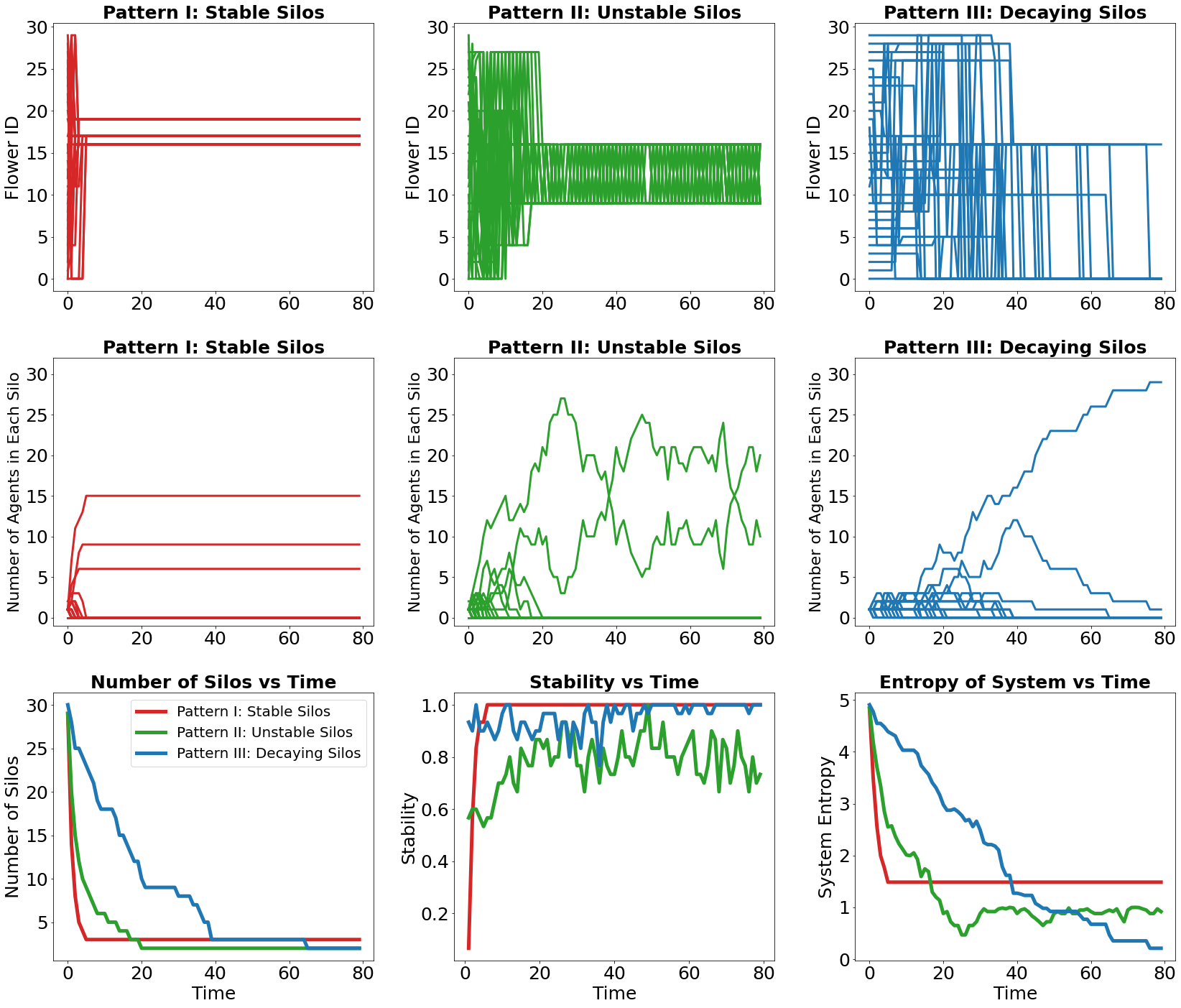}
\caption{Three patterns of silos: Stable (Pattern I), Unstable (Pattern II), and Decaying (Pattern III), coming from different pairs of ($p,k$). 
System classification occurs at $ T = 80 $. 
\textbf{Top.} Example systems for the three of the observed types of patterns. 
Each line represents an agent.
\textbf{Middle.} The dynamics of the number of agents in each silo for each of the example systems in the top row.
Each line represents a silo.
\textbf{Bottom.} The metrics used to classify a system for each of the example systems in the top row.
Systems with stable silos have multiple non-empty silos with unchanging members; systems with unstable silos have multiple non-empty silos with interchanging members; systems with decaying silos have recently experienced the observed minimum entropy.
}
\label{fig: 3patterns}
\end{figure*}

\item \textbf{(Interaction Step)} At each time $t>0$, we update each agent's database based on their interaction with another agent.
The agents in which agent $ A_{i} $ can communicate with is determined by its $ k $-nearest neighbors (says $ D^{(t)} $).
In particular, $ A_{i} $ will interact with exactly one of its nearest neighbors.
Given that $ A_{j} $ is in $ A_{i}$'s nearest neighbors, the probability that $ A_{i} $ interacts with $ A_{j} $ is $ 1 / k $.
The type of interaction that $ A_{i} $ has with $ A_{j} $ is determined as follows:
\begin{itemize}
    \item With probability $ p $ (constant for all agents), agent $ A_{j} $ mirrors $ A_{i} $.
    Functionally, this means $ A_{i} $ ``updates" itself with its own answer $ B_{i}^{(t)} $.
    This update function serves to model agent $ A_{j} $'s social alignment to $ A_{i} $  \cite{Hasson_Frith_2016a}. 
    \item With probability $1-p$, $ A_{j} $ does not mirror $ A_{i} $ and, instead, responds with $ B_{j}^{(t)} $.
    $ A_{i} $ is then updated with $ B_{j}^{(t)} $.
    This update function is a model of basic information exchange wherein individuals communicate without copying behavior (or, do not align) before responding.
\end{itemize}
\end{itemize}

An illustration of system dynamics is provided in Figure \ref{fig:system_illustration}. Note that the evolution of the described system depends on the probability of interacting with a mirroring agent ($ p $), and the number of nearest neighbors an agent is able to interact with ($ k $; sometimes referred to as the agents' ``range of communication").
We measure system evolution by clustering the agents' responses based on the flower species that they say is the prettiest. 
To extract the exact species names from agent responses, we define a comprehensive list of possible flower names and use a Python program to find the names in sentences. 
For analytical simplicity, we assign each flower a Flower ID 
and use it to designate an agent's cluster (or ``silo") membership.
Silo membership can change as a function of $ t $.
We note that the embeddings of the descriptions capture the similarities across flower species that we do not fully encode in the Flower IDs.

We let $ c_{i}^{(t)} $ denote the silo in which $ A_{i} $ is a member of at time $ t $ and $ \omega_{c}^{(t)} $ denote the number of agents in silo $ c $ at time $ t $.
We define the stability of the system at time $ t $, $ S^{(t)} $, to be the proportion of agents where $ c_{i}^{(t)} = c_{i}^{(t-1)} $:
\begin{equation*}
    S^{(t)} := \frac{1}{n} \sum_{i=1}^{n} \mathbbm{1}\{c_{i}^{(t)} = c_{i}^{(t-1)} \}.
\end{equation*}
We similarly let $ E^{(t)} $ be the entropy of the system at time $ t $ \citep{manning2008}, 
\begin{equation*}
    E^{(t)} := -\sum_c \frac{\omega_c^{(t)}}{n}\log_2\frac{\omega_c^{(t)}}{n}.
\end{equation*}
We analyze the evolution of systems using $ S^{(t)} $ and $ E^{(t)} $ as a function of $ p $ and $ k $ below.
\section{Results}
\subsection{Observed silo patterns\label{sec:silo_patterns}}

\begin{figure*}[t]
\centering
\includegraphics[width=11.4cm]{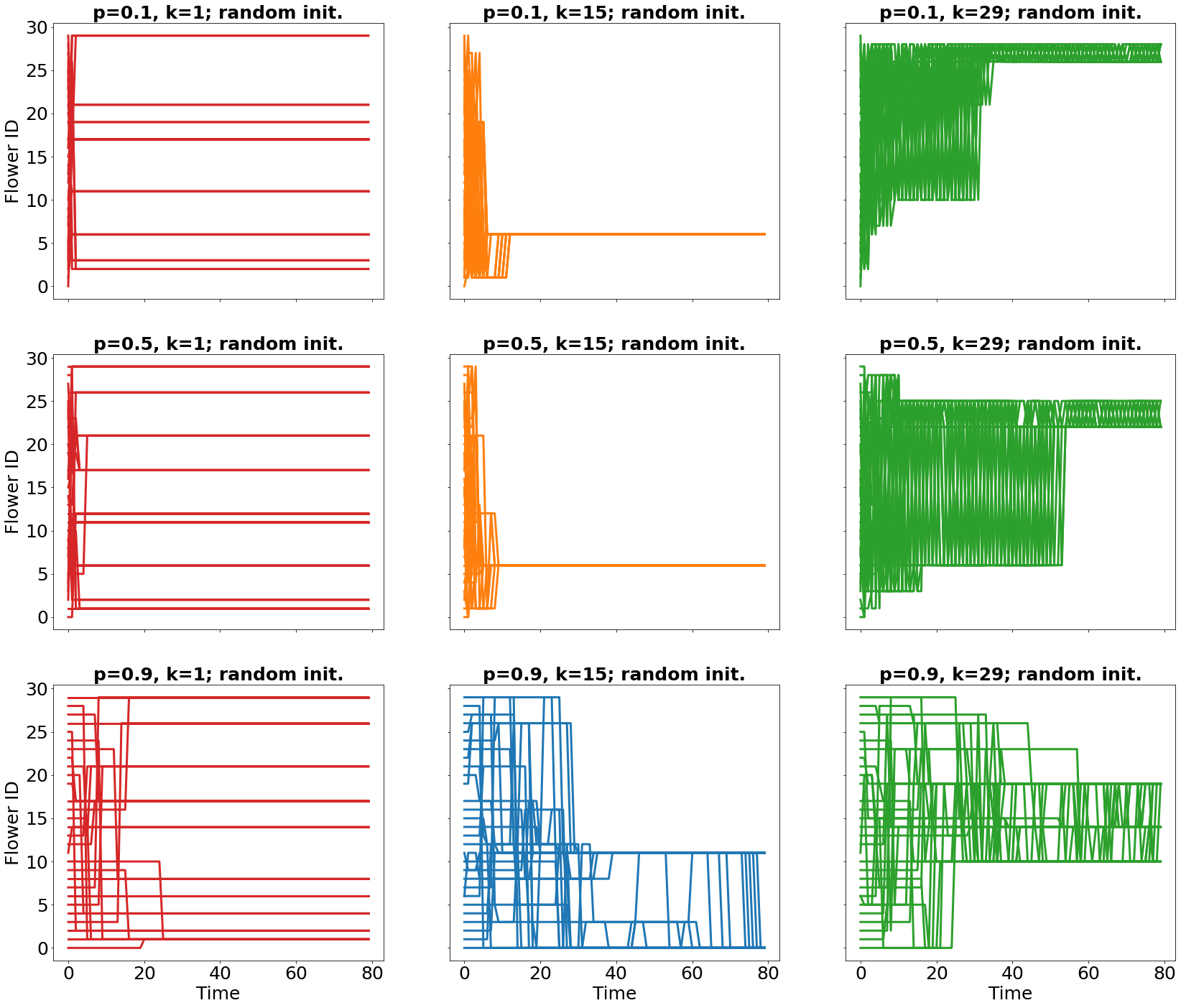}
\caption{Example systems of $ n = 30 $ with different agent behaviors. 
Graph color (red/blue/green/orange) indicates the pattern of stable/decaying/unstable/one silo(s) at $ T = 80 $.
Small range of communication for each agent ($ k $) appears to prohibit global alignment.
Large likelihood of mirroring ($ p $) delays global alignment.
We investigate these relationships further in Figures \ref{fig: count_vs_k} and \ref{fig: count_vs_p}.}
\label{fig: 9flowerIDs}
\end{figure*}

Before discussing the effects of $ p $ and $ k $ on the evolution of the system, we first highlight three patterns of silos that we observed from our experiment: I) ``Stable silos", II) ``Unstable silos", and III) ``Decaying silos". We characterize Patterns I-III via the number of silos at $ t $, $ S^{(t)} $, and  $E^{(t)} $:
\begin{enumerate}[I.]
    \item A system has stable silos if, after some $ t^{*} > 0 $, the number of silos is constant, $ E^{(t)} $ is constant, and $ S^{(t)} = 1 $. 
    A system with stable silos is in a steady state.
    \item A system has unstable silos if, after some $ t^{*} > 0 $, the number of silos is constant, $ E^{(t)} $ is approximately constant, and $ S^{(t)} < 1 $. 
    A system with unstable silos is in a steady state.
    \item A system has decaying silos if
    $T-\min(\underset{t}{\mathrm{argmin}} \;E^{(t)})<m$ for a suitably chosen $ m $. 
    That is, the system is decaying if the first time the system entropy achieves its minimum is within $ m $ steps from the end of the system's evolution. 
    In our experiment $T = 80 $ and we chose $ m = T / 10 = 8 $.
    A system with decaying silos is not in a steady state.
\end{enumerate}
We also observed systems that converge into a single silo. 
We refer to these systems as ``One silo" systems and note that they are in a steady state.
Systems observed to have Pattern III will eventually decay into a system with stable silos, unstable silos, or a single silo.
Further, while different subsystems within a system may exhibit different (un)stability or decay, we classify the systems using the number of silos, entropy, and stability of the entire system.

We show an example of Patterns I-III in the top row of Figure \ref{fig: 3patterns}.
Each line corresponds to an agent's silo evolution up to $ T=80 $.
The leftmost figure shows a system exhibiting Pattern I, where the Flower ID for each agent stabilizes and remains the same.
Conversely, the center figure shows a Pattern II system where some agents in a Pattern II system oscillate between at least two silos for $ t > 20 $.
The rightmost figure shows an example of a decaying system: one of the silos is slowly absorbing the others.

The second row of figures show how the number of agents in a silo evolves over time for each of the example systems in the first row. 
Each line in a given figure corresponds to a silo.
In the system with stable silos, the number of agents stabilizes quickly.
In the system with unstable silos, the number of members in two of the silos oscillate, as agents ``jump" from one silo to the other and back.
Neither of the oscillating silos appear primed for substantial decay or growth in the long-run -- a hallmark characteristic of the group dynamics in a system with unstable silos.
Lastly, the system with decaying silos has a silo that is nearly always growing while the other silos do not attract more members after $ t \approx 40 $. 

The last row of figures shows the pattern-defining metrics associated with each highlighted system: number of silos (left), stability (center), and entropy of system (right).

\subsection{The effects of p and k}
\label{sec:the_effect_of_p_and_k}
Figure \ref{fig: 9flowerIDs} shows the evolution of systems of $ n = 30 $ agents for nine different settings of agent behaviors (i.e., $p$ and $k$). 
Overall, when the agents' range of communication is small we observe stable silos; when $ k / n \approx 0.5 $ and $ p $ is small we observe one silo; and when $k\xrightarrow{}n$ we observe unstable or decaying silos. 
Further, when the likelihood of interacting with a mirroring agent is small, systems converge quickly; when $ p $ is large, systems stay in transient states for longer.

\begin{figure}
\centering
\includegraphics[width=\linewidth]{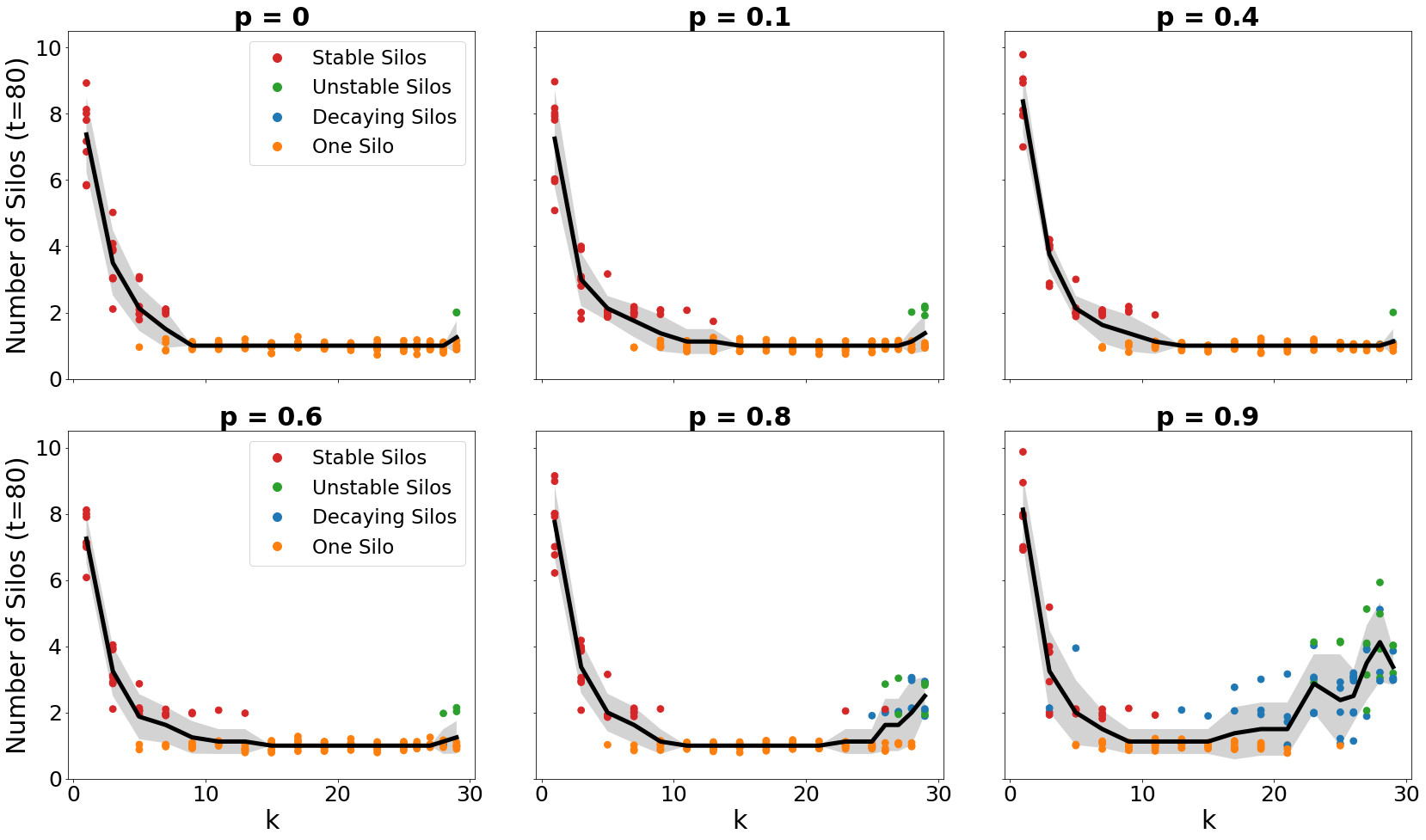}
\caption{Number of Silos vs. $k$ (the range of communication) for several values of $p$ (the likelihood of an agent interacting with a mirroring agent). For each setting of agent behavior we include the number of silos observed at $ T = 80 $ for $ 8 $ random agent initializations. 
Dot color corresponds to system type at $ T = 80 $.  
The black line is the average number of silos with the shaded area representing  $ \pm ~ 3 $ standard errors.
Up to a point, increasing $ k $ encourages global alignment. 
When $ k $ is large the system is more likely to contain multiple silos at $ T = 80 $.}
\label{fig: count_vs_k}
\end{figure}

\begin{figure}
\centering
\includegraphics[width=\linewidth]{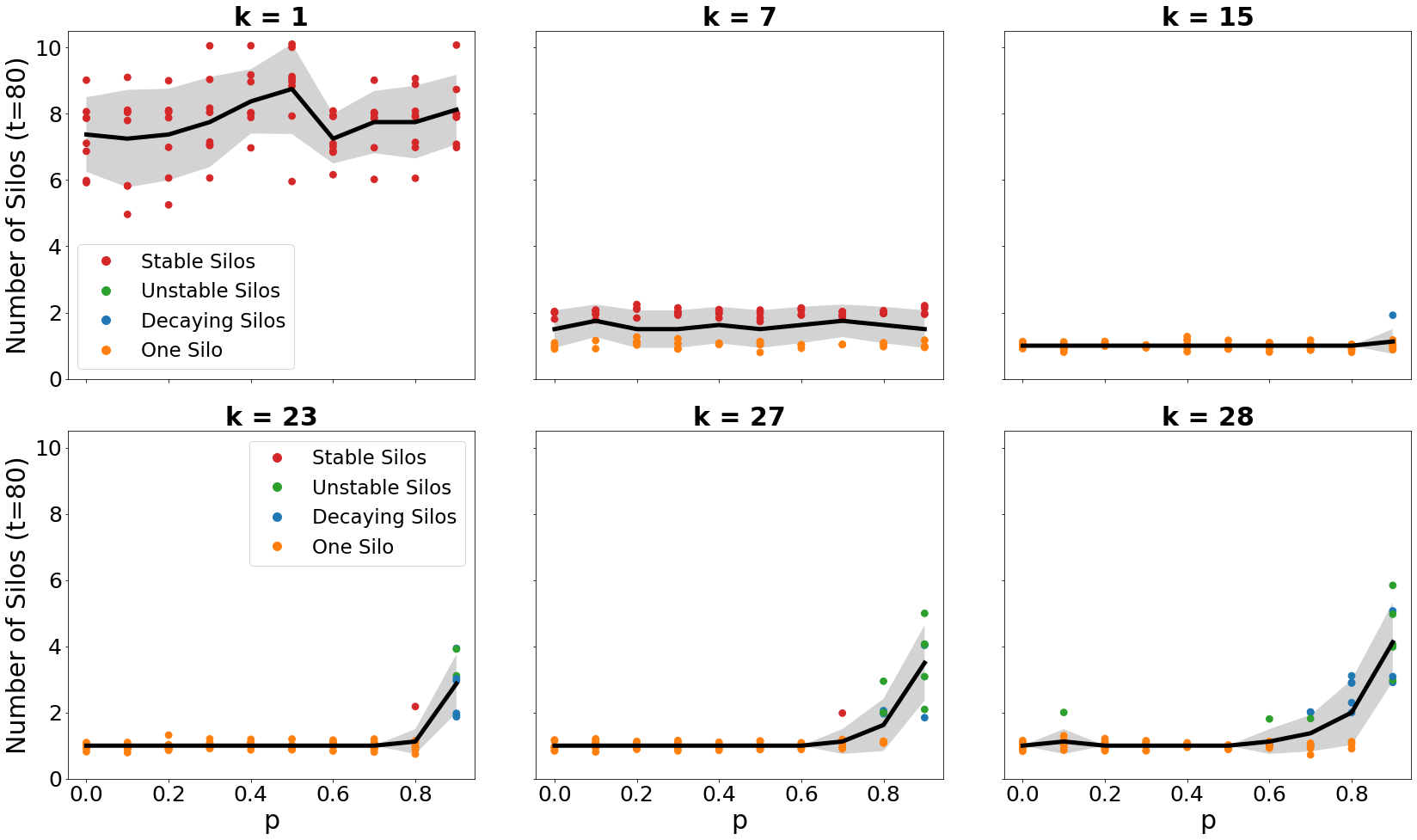}
\caption{Number of Silos vs. $p$ (the likelihood of an agent interacting with a mirroring agent) for several values of $k$ (the range of communication). For each setting of agent behavior we include the number of silos observed at $ T = 80 $ for $ 8 $ random agent initializations. 
Dot color corresponds to system type at $ T = 80 $.  
The black line is the average number of silos with the shaded area representing $ \pm ~ 3 $ standard errors.
Increasing $ p $ decreases the likelihood of global consensus across all $ k $.}
\label{fig: count_vs_p}
\end{figure}

We explore the relationship between system type and number of silos at $ T = 80 $ in more detail in Figures \ref{fig: count_vs_k} and \ref{fig: count_vs_p}, where each dot corresponds to an entire system.
In particular, for each setting of agent behavior we consider $ 8 $ different random agent initializations and label each system according to the observed system behavior up to $ T = 80 $. 
We include a line representing the average number of observed silos at $ T = 80 $ to help emphasize the effects of $ p $ and $ k $.

Figure \ref{fig: count_vs_k} shows the relationship of silo count and $k$ for various $p$. As observed in Figure \ref{fig: 9flowerIDs}, when $ k $ is small the systems typically contain stable silos.
For $ k / n \approx 0.5 $ systems typically contain a single silo.
For $ k / n > 0.5 $, systems are more likely to contain unstable silos or decaying silos as $ k $ increases, though this effect is dependent on $ p $.

Our results suggest that the range of communication of the agents acts as a moderating pressure on the number of supportable silos: when the range is partially restrictive the agents are able to form a consensus, when the range is too restrictive or too loose the system experiences population polarization or multi-silo instability, respectively. 
As a consequence, fractured systems will remain fractured in settings where agents have a restricted communication range and multiple unstable silos are likely to persist in settings where agents have a nearly unbounded range of communication.

In Figure \ref{fig: count_vs_p}, we investigate the relationship of silo count and $p$ for various $ k $. 
For small $ k $, $ p $ does not appear to have a large impact on the number of silos observed at $ T = 80 $. 
For $ k / n $ close to one, however, $ p $ has a clear effect on the number of silos.
As $ p $ increases the expected number of silos increases.
Further, these silos are typically unstable or decaying.

The splintering of high-$p$, high-$k$ populations is primarily due to the increase in time which a high-$p$ population requires to converge. 
In particular, when the majority of interactions are instances of a mirroring interaction, there is little opportunity for actual information to be shared. 
Thus, agents are more likely to remain isolated in their own perspective bubble for a longer amount of time -- with only the off-chance that basic interactions serving to collapse perspective bubbles into multi-agent silos -- resulting in few instances where a consensus across the entire population is reached by $ T = 80 $. 
We note that when $ p = 1 $, the number of silos is equal to the number of flower IDs present in the initial conditions, as the opportunity for an informative interaction has been completely replaced by mirroring interactions. 
 
\section{Discussion}

In Section 2.\ref{sec:silo_patterns} we introduced a system classification based on the observed silo behavior in a system of interacting agents.
In Section 2.\ref{sec:the_effect_of_p_and_k} we discussed the effects of the range of agent communication and of the likelihood of interacting with a mirroring agent on the number and type of silos observed at $ T = 80 $.

Our results show that increasing the rates of social mirroring can have substantial impact on the group dynamics of a system of interacting agents.
The effect is particularly pronounced in settings where the range of communication for each agent is suitably large.
For settings where agents have more restricted communication, high rates of social mirroring have a minimal impact on the degree and type of group formation.
We interpret these results as due to a perspective-filtering effect inherent to systems with local communication.
Indeed, local communication settings are essentially high-mirroring settings since agents only interact with others similar to them.
When $ p $ is large we observe multiple groups at $ T = 80 $, though the type of groups that arise depends on $ k $.

The observed effect of $ p $ and $ k $ on the dynamics of the system has notable parallels with known behaviors of observed in social groups.
For example, research investigating the role of information overexposure in polarization has shown that populations exposed to increasing amounts of of information tend towards being more polarized \cite{polarization, polarisation}.
We observe similar behaviors in our simulated systems, particularly as the trend of group count decreasing as $k$ increases reverses near the global communication limit, resulting in multiple unstable and/or decaying groups -- closely matching the observed behaviors in in populations inundated with vast amounts of information \cite{polarization, polarisation}.

Similarly, \cite{family} and \cite{echo} discuss potential causes of the formation of physical and online echo chambers.
Essentially, members in family-sized groups reinforce the opinions of fellow members.
For physical groups, opinion is fortified based on pre-existing geographical proximity \cite{family}.
For online echo chambers, members self-select based on their pre-existing opinions and thus are only ever subjected already-aligned people \cite{echo}.
Different parameter settings of our simulated system allow us to understand the effect of these behaviors when applied to every agent in the system --
small $ k $ models opinions reinforced by geographical communication constraints and large $ p $ models opinions reinforced after self-selecting for interactions with agents with similar opinions. 
As seen in Figures \ref{fig: count_vs_k} \& \ref{fig: count_vs_p} these settings lead to many small stable groups that are unable to achieve population consensus due to their isolation from alternative perspectives.

Finally, recent studies explore the links between social media usage and value mirroring \cite{social_media} and detail the increasing rates of mirroring via self-censorship \cite{selfcensor}.
These studies suggest that group-membership prediction is an increasingly difficult task as people are unwilling to publicly break from perceived policy and social norms -- potentially providing a reason why polling accuracy has not increased despite the availability of more data on the electorate \citep{jennings2018election}.
Indeed, the results from our simulated systems suggest that high rates of social mirroring could cause poor estimation of support for a particular candidate or piece of legislation since systems with large $ p $ are more likely to have multiple unstable or decaying silos at $ T = 80 $. 

Our research demonstrates that social mirroring serves as a modifier for the pre-existing trend dictated by the range of communication in a population. 
Overall, the range of interaction serves as a predictive variable with respect to the state of silos in a population -- whether they are stable, unstable, or decaying, as well as the silo count. The frequency of social mirroring, however, distorts this base trend by exaggerating it as mirroring increases in frequency. A population experiencing global communication will likely have multiple unstable groups regardless of $p$, but the magnitude of silo splintering and the severity of instability will be shaped significantly by $p$.
\section{Limitations}
Our framework enables simple and interpretable investigation of the effect of the range of agent communication and the likelihood of interacting with a mirroring agent on group dynamics and social alignment.
With that said, there are many interesting directions for follow-on work.

\begin{figure}[t]
\centering
\includegraphics[width=0.9\linewidth]{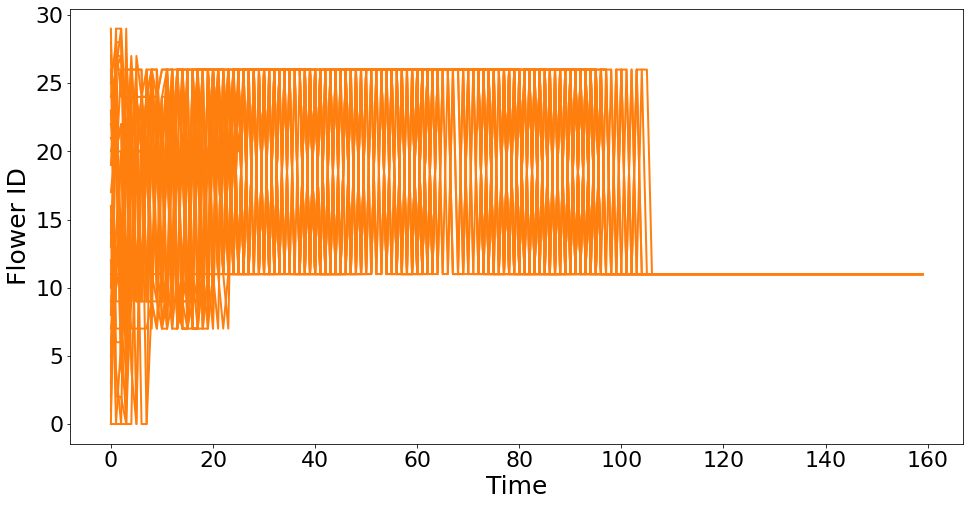}
\caption{Example system with $p=0.2$, $k=29$, $T=160$. The system has unstable silos at $ T = 80 $ and a single silo for $ T > 105 $ -- indicating that longitudinal analysis may provide additional insights into system behavior.}
\label{fig: T160}
\end{figure}

For example, we limited our analysis to systems with $ n = 30 $ agents.
While $ 30 $ agents is large enough for relatively complicated group dynamics to emerge, simulating larger systems will allow us to draw parallels to the dynamics observed in human systems more readily. 
Similarly, we focused on system classification at $ T = 80 $. 
The classification of a system will depend on the choice of $ T $, as seen in Figure \ref{fig: T160} where we observe unstable silos at $ t = 80 $ and a single silo for $ t > 110 $. 
Extending our work to the longitudinal effects (i.e., across various $ T $) of system hyperparameters may be necessary to fully understand group dynamics and final consensus behavior. 

In addition, our analysis is focused on systems with a fixed ($p,k$) across time.
This is a simplification that we make to present digestible results, though we recognize that it is an abstraction that may not properly reflect real-world interactions, as individuals adjust their communication size and level of mirroring based on multiple factors.
Extending our analysis to a more dynamic setting may be very valuable.

Further, our framework can be used to study the effect other social phenomena -- such as miscommunication -- on alignment by adjusting the update mechanism.
Recall that in our analysis the database for each agent is updated to perfectly reflect its most recent interaction.
We could introduce variably sized perturbations of the response from the most recent interaction to model different levels of agent miscommunication.
Such analysis would provide insight into how repeated exposure to information may impact alignment.

We could similarly increase the complexity of the agent's databases and retrieval mechanisms to better model human behavior. 
For example, the generative agents studied in \cite{park2023generative} store, process, and manage a history of complicated interactions with their environment.
While there are benefits to the simple agents that we study, increasing the complexity of the agents while maintaining the simple interaction mechanisms studied herein may improve the generalizability of our simulations to sociological settings.
We note that increasing the complexity of the agents will require using general black-box techniques \citep{duderstadt2024, helm2024tracking, acharyya2024, helm2024} to study agent, group, and system dynamics.

Lastly, we note that the behavior of the systems that we studied can be approximated with a system of interacting Gaussian Mixture Models (GMMs). 
On one hand, replacing the current agents with GMMs may decrease the generalizability of the results to human social systems.
On the other hand, it could make the simulations more computationally efficient and more mathematically tractable. 
The cost effectiveness of a system of interacting GMMs, even relative to the simple agents studied herein, could enable the analyses of systems with an otherwise impossible number of agents and time steps.
The tractability of the GMM systems could enable theoretical analysis of the group dynamics and limiting behavior in terms of $ E^{(t)} $ and $ S^{(t)} $. 


\bibliography{citation}

\end{document}